# Robustness and modular design of the *Drosophila* segment polarity network

Wenzhe Ma[1,2], Luhua Lai[1,2], Qi Ouyang[1,3,*] and Chao Tang[1,4,*]

1. Center for Theoretical Biology, Peking University, Beijing 100871, China
2. Department of Chemistry and Molecular Engineering, Peking University, Beijing 100871, China
3. Department of Physics, Peking University, Beijing 100871, China
4. Departments of Biopharmaceutical Sciences and Biochemistry and Biophysics, University of California, San Francisco, CA 94143, USA

* Corresponding authors: Q Ouyang, Department of Physics, Peking University, Beijing 100871, China. Tel.: +86 10 6275 6943; Fax: +86 10 6275 9041; E-mail: qi@pku.edu.cn or C Tang, Department of Biopharmaceutical Sciences, University of California, San Francisco, CA 94143-2540, USA. Tel.: +1 415 514 4414; Fax: +1 415 514 4797; E-mail: chao.tang@ucsf.edu

**Abstract**

Biomolecular networks have to perform their functions robustly. A robust function may have preferences in the topological structures of the underlying network. We carried out an exhaustive computational analysis on network topologies in relation to a patterning function in *Drosophila* embryogenesis. We found that while the vast majority of topologies can either not perform the required function or only do so very fragilely, a small fraction of topologies emerges as particularly robust for the function. The topology adopted by *Drosophila*, that of the segment polarity network, is a top ranking one among all topologies with no direct autoregulation. Furthermore, we found that all robust topologies are modular—each being a combination of three kinds of modules. These modules can be traced back to three sub-functions of the patterning function and their combinations provide a combinatorial variability for the robust topologies. Our results suggest that the requirement of functional robustness drastically reduces the choices of viable topology to a limited set of modular combinations among which nature optimizes its choice under evolutionary and other biological constraints.

# Introduction

Biological systems are evolved to function robustly under complex and changing environments (Waddington, 1957). At the cellular level, the interactions of genes and proteins define biomolecular networks that reliably execute various functions despite fluctuations and perturbations. Functional robustness as a systems property may have preferences in and constraints on the wiring diagram of the underlying networks (Barkai & Leibler, 1997; Li et al, 2004; El-Samad et al, 2005; Wagner, 2005). It has been demonstrated in a computational study that a robust oscillator has a strong preference on certain type of the network topology (Wagner, 2005). Preferred network motifs in biological networks were identified (Milo et al, 2002) and were attributed to their robust dynamical properties (Prill et al, 2005). It was argued through a comparative study of a few networks that a bacteria signaling network is optimally designed for its function (Kollmann et al, 2005). To clearly lay out the relationship between the functional robustness and the topological constraints, we carry out an exhaustive computational analysis on the network topologies that perform the same patterning function as the segmentation polarity gene network in *Drosophila* (Matizez Arias, 1993; DiNardo et al, 1994; Perrimon, 1994). We found that only a small fraction of topologies can perform this patterning function robustly. This information can be used in combination with mutant phenotypes to discriminate biological models. We show that the topology of the *Drosophila* network is among this small group of robust topologies and is optimized within certain biological constraints. We further found that all robust topologies can be classified into families of core topologies. Each family is a particular combination of three kinds of network modules which originate from the three sub-functions of the patterning function. We argue that the modular combinations also facilitate flexibility and evolvability in this case.

The segmentation process in the embryogenesis of the fruit fly *Drosophila* is characterized by a sequential cascade of gene expression, with the protein levels of one stage acting as the positional cues for the next (Wolpert, 2002). The successive transient expression of the maternal, the gap and the pair-rule genes divide the embryo into an ever finer pattern. After cellularization, the segment polarity genes stabilize the pattern, setting up the boundaries between the parasegments and providing positional "readouts" for further development (Matizez Arias, 1993; DiNardo et al, 1994; Perrimon, 1994). We are concerned here only with the network that is in action during the extended and the segmented germband stage which is characterized by the interdependency of the expression of *en* and *wg* (DiNardo et al, 1994; Perrimon, 1994), and we focus on its function of stabilizing a periodic pattern of sharp boundaries defined by the *en*- and the *wg*-expressing cells (Vincent & O'Ferrell, 1992). As depicted in Figure 1, the core network in *Drosophila* consists of the hedgehog (Hh) (Lum & Beachy, 2004) and the wingless (Wg) (Klingensmith & Nusse, 1994) signal transduction pathways. Previous studies demonstrated that this network is a very robust patterning module. Differential equation models of the network can stabilize and maintain the required patterns of *en* and *wg* expression with a remarkable tolerance to parameter changes (von Dassow et al, 2000; von Dassow & Odell, 2002; Ingolia, 2004). A simple Boolean model was shown to capture the main feature of the network's dynamics (Albert & Othmer, 2003). These findings have led to the hypothesis that the segment polarity gene network is a very robust developmental module that is adopted in a wide range of developmental programs (von Dassow et al, 2000). Indeed, the striped expression patterns of the segment polarity genes in the segmented germband stage are remarkably conserved among all insects, perhaps among all arthropods (Peel et al, 2005). On the other hand, it was argued that the conservation of this gene network is not due to robustness but rather to pleiotropy (high connectivity with other modules/networks) (Sander, 1983; Raff, 1996; Galis et al, 2002). Pleiotropic effects may constrain the network's evolution, "freezing" its topology early on during evolution and making it conserved among developmental programs that later diverged. In this study, we investigate the relationship between the functional robustness and the network's topology. Specifically, we ask (1) How many network topologies can perform the given patterning function and how many can do so robustly? (2) Can a robust topology also satisfy certain topological constraints imposed by, e.g. pleiotropic effects, and if so how is this achieved? (3) Where does the *Drosophila* network stand in this analysis? (4) Are there any organization principles emerging from the robust topologies for the given function?

## Results

### Coarse-graining the biological network

Instead of analyzing the full biological network, we focus on its core topology. The core topology is derived from the full network and is the minimal set of nodes and links that represent the underlying topology of the full network. This reduction in degrees of freedom enables us to perform a much more comprehensive computational and theoretical analysis and at the same time to preserve key functional properties. The topology of the *Drosophila* segment polarity network can be represented by a network of three nodes. The network represented in Figure 1A can be simplified into the topology of Figure 1B. Since we are mainly concerned with the steady-state behavior, certain "intermediate steps" in the network can be combined. First, we combine the mRNA node with its corresponding protein node if there is no posttranscriptional regulation for the mRNA, because the time delay between the mRNA and the protein production does not play any role in our steady state analysis. We then combine the node hh/Hh with en/En, because the expression of *hh* depends solely on En. The expression pattern of these two genes, *hh* and *en*, are highly correlated at this stage of the development (Tabata et al, 1992). We thus use a single node "E" in Figure 1B representing the four nodes, *en*, En, *hh*, Hh in Figure 1A. Extra-cellular Hh signaling activates *wg* by regulating the expression of Ci and Cn, which are parts of the Hh signal transduction pathway. Both Ci and Cn are the products of the gene *ci*. In the absence of Hh signaling, Ci goes through a process of proteolysis and the remaining fragment functions as a repressor, Cn. The Hh signaling blocks the proteolysis of Ci, resulting in the accumulation of Ci in the nucleus, which acts as a transcriptional activator for *wg* (Lum & Beachy, 2004; Alexandre et al, 1996). Thus, Ci and Cn function like a transcriptional switch in response to the Hh signaling. This regulation is simplified as a direct (intercellular) link from "E" to "W" in the coarse grained topology (Figure 1B). The repression of *ci* by En in Figure 1A is represented in Figure 1B as "E" repressing "W" since the function of *ci* is to control the expression of *wg*. The simplified model Figure 1B has similar dynamic properties with the more detailed model Figure 1A; in particular they can both stabilize the wild type pattern and sharpen the parasegment boundary.

In coarse-graining the network of Figure 1C to that of Figure 1D, the two negative regulations, from Slp to *mid* and from Mid to *wg*, are replaced with a direct positive regulation from "S" to "W". Again, the simplified model Figure 1D has similar dynamic properties and patterning function with the full model Figure 1C.

**Enumerating 3-node networks**

We then proceed to enumerate topologies of 3-node networks with intra- and intercellular interactions. Every node may regulate itself and the other two nodes, both intracellularly and intercellularly, resulting in 3×3×2=18 directed links. Each link has three possibilities: the regulation can be positive, negative, or absent. So the total number of possible topologies for the 3-node network is $3^{18}$= 387,420,489. Enumerating all of them is beyond our computational power. Thus we make the following restrictions on the topology: only two out of the three nodes can possibly go outside of the cell to signal. This restriction reduces the total number of topologies to $3^{15}$=14,348,907, all of which we enumerate. For each topology, we use a model of ordinary differential equations to quantitatively assess its ability to perform the required function, which is to stabilize the pattern of Figure 1F given the initial condition of Figure 1E (Methods). The functional robustness of a topology is measured by the quantity Q=[the fraction of the parameter space that can perform the function] (von Dassow et al, 2000; Ingolia, 2004). We estimate Q by randomly sampling the parameter space: Q≈m/n, where n is the number of the random parameter sets used in the sampling and m the number of those sets that can perform the function. We first sampled each and every topology with n=100 random parameter sets. We found that about 1% of the topologies can perform the function with at least one of the 100 parameter sets (m>0). However, their Q values differ drastically. As shown in Figure 2, the distribution of the Q values is much skewed among the 1% population of the topologies—while the majorities have very small Q values there is a long tail in the distribution.

**Biological network**

The topology (Figure 1B) of the network constructed in previous studies (von Dassow & Odell, 2002; Ingolia, 2004) (Figure 1A) scored very high but is not the top ranking one. However, there may be some biological constraints on the selection of topologies. Indeed, a group of topologies consisting of only two nodes (with the "S" node left unlinked) come close to the top (see Figure 3A), suggesting that if *Drosophila* were only presented with the function defined in our study the best design would be to just use two mutually activating signaling pathways ("E" and "W") and nothing else. But both the Hh and the Wg signaling pathways are utilized in at least several other functions besides stabilizing the parasegment boundaries (Galis et al, 2002), which may impose pleiotropic constraints on the topology of the networks that utilize these pathways. In general, these constraints may be hard to decipher. Here we simply note that there is no sound biological evidence for any *direct* positive autoregulation loops on the two signaling pathways and on the *slp* genes. If we exclude topologies with any direct positive autoregulation on the "E" and the "S" nodes, Figure 1B stands up as the most robust topology (Q=0.47). In this topology, there is still a direct autoregulation loop on "W", which originates from the Wg -> *wg* autoregulation in Figure 1A. This autoregulation has no basis in biological evidence, but was added by previous authors to ensure the correct patterning of the model—without this added link, their models cannot reproduce the correct biological pattern (von Dassow et al, 2000; von Dassow & Odell, 2002; Ingolia, 2004). We ask that if we do not add this autoregulation whether we can identify a robust topology that has biological evidence for every link. There are 8 topologies with Q>0.1 that have no direct autoregulation on any of the three nodes. A top ranking one (Q=0.36) is shown in Figure 1D. Instead of a direct autoregulation on "W", this topology accomplishes the positive feedback indirectly through the node "S". This would suggest that the Wg signaling pathway regulates the *slp* gene whose product in turn regulates *wg*. Indeed, there is ample biological evidence for these regulations (Lee & Frasch, 2000; Buescher et al, 2004), suggesting a biological network of Figure 1C. The role of *slp* in regulating *wg* was also discussed in previous computation models (Meir et al, 2002; Albert & Othmer, 2003).

To further determine which of the two topologies, Figure 1B or D, is closer to the true biological one, we subject both to the mutant test. We model two kinds of mutants corresponding to perturbations in the two signaling pathways and compare the computed phenotypes with the experimental observations. The first is the *zw3* (a protein kinase in Wg signaling pathway) mutant—the mutation results in a ubiquitous Wg signaling with a phenotype of an expanded *en*-expressing region ended by an ectopic *wg*-expressing stripe (Figure 1G) (Siegfried et al, 1994). The second is the mutation of the Hh receptor *patched* (*ptc*) which results in a ubiquitous Hh signaling and has a phenotype of an expanded *wg*-expressing region ended by an ectopic *en*-expressing stripe (Figure 1H) (DiNardo et al, 1988). We found that while both topologies, Figure 1B and D, can produce the wild type patterning robustly, only the network of Figure 1D can also produce the two mutant phenotypes. Specifically, for Figure 1D about 1/3 of the parameter sets that produced the wild type pattern can also produce the two mutant patterns. For Figure 1B, none of the parameter sets that produced the wild type pattern can also produce either of the two mutant patterns. We also used the more detailed models Figure 1A and C to carry out the mutant test and obtained similar results (see details in Supplementary Information). This suggests that the network of Figure 1C (and its corresponding topology of Figure 1D) is a better model for the *Drosophila* network than that of Figure 1A (and Figure 1B).

## 2-node topologies

In our enumeration study of the 3-node topologies, there are some 2-node topologies (with the "S" node unlinked) scored very high. This indicates that the simplest "irreducible" topology for the required patterning function consists of only two nodes and that it would be instructive to study 2-node topologies. There are 45 2-node topologies with Q>0.1. A close examination of these topologies revealed that all of them come from 4 core topologies, which we call skeletons (Figure 3A). In other words, the 45 topologies can be classified into 4 families. In each family, all the members come from a skeleton by adding extra links to the skeleton. These links are either "neutral" (have no effect on the Q-value) or "bad" (will reduce the Q-value). The number of neutral links a skeleton can accommodate and the number of bad links it can tolerate (so that the reduced Q value is still large than 0.1) depend on the structure and the robustness of the skeleton. As shown in Figure 3A, the first skeleton can accommodate and tolerate combinations of 2 neutral links and 4 bad links, while the fourth skeleton can accommodate or tolerate none. Furthermore, the 4 skeletons all contain the following three topological features: positive feedback on "E" (either intra- or intercellularly), positive feedback on "W" (either intra- or intercellularly), and intercellular mutual activation between "E" and "W". These three topological features can be traced back to three sub-functions which the required patterning function can be decomposed into. Note that cells adjacent to an "E"-expressing cell can have two different fates: expressing "W" or none (Figure 1F). So the network should be bistable in "W". Similarly, cells adjacent to "W" can express either "E" or none, implying bistability in "E". Thus the positive feedback loops on "E" and on "W" follow the functional requirement of bistability on "E" and "W" (Ingolia, 2004). The mutual intercellular activation between "E" and "W" arises from the functional requirement of maintaining a sharp patterning boundary. In order to sharpen a wide boundary (Figure 1E), it is necessary to have "E" expressed only right next to a "W" cell, and vice versa, leading to the interdependency of "E" and "W" in the network topology. Therefore, the three functional requirements lead to the three kinds of topological features, or modules. The combination of the three kinds of modules, with one from each kind, results in the 4 skeletons of the robust topologies (Figure 3A). Note that for the second, the third and the fourth skeletons in Figure 3A, there are necessary repressive links (which are neutral in the first skeleton) in addition to the three modules. When the positive feedback module is intercellular, it is necessary to have an intracellular repression on the node to prevent the "E" and "W" being expressed in the same cell causing further blurring of the boundary (see details in Supplementary Information). Also note that some bad links are just redundant modules, e.g. the intercellular auto-activation of E or W in the first skeleton.

## Families of 3-node topologies

Having identified the three essential kinds of modules for the patterning function and the rules of their combination in 2-node topologies, we turn our attention to the robust 3-node topologies and ask if similar organization principles exist there. With one extra node "S", there are multiple new ways to form each kind of modules (Figure 3B). (Note that for the E and W modules the positive feedback does not have to act on E/W directly. If E/W is dependent on S, positive feedback on S is also a viable choice. Also note that since we have excluded from our enumeration the intercellular regulation from S, there are no modules with this regulation.) We then checked all 3-node topologies with $Q>0.1$ to see if they contained these modules. Intriguingly, every topology in this pool (37,580 of them) contains at least one module of each kind. Therefore, it is a necessary condition for a robust topology to include at least one module of each kind. On the other hand, the reverse is not true. From the modules in Figure 3B, one can form 108 combinations that include one and only one module of each kind and that have no conflicting regulations (see details in Supplementary Information). Only 44 of them are robust enough to be the skeletons of networks with Q>0.1. In other words, we found that all topologies with Q>0.1 can be classified into 44 distinct families corresponding to 44 modular combinations (skeletons). In most families, the Q value of the skeleton is either the highest or close to the highest in the family, implying that other members in the family have extra non-beneficial (neutral and bad) links compared to the skeleton. There are a few cases where the skeleton's Q value is not close to the top within the family, implying that some extra links in addition to the modular combination are beneficial.

As shown in Figure 4A, the family size roughly scales exponentially with the skeleton's Q value. This means that the larger the skeleton's Q value, the more non-beneficial links it can accommodate and tolerate. The exponential dependence of the family size on the Q value suggests family members as some kind of combinatorial additions to the core topology, although in general the effects of additions of links to the core may be correlated. While the non-beneficial links do not improve the Q values, they may facilitate variability and plasticity that can be useful in adapting to new environments and functional tasks (Schuster et al, 1994). We found that certain neutral links and redundant modules are beneficial when the system is faced with noisy initial conditions (see details in Supplementary Information). The modular organization of the skeletons suggests that their Q values might be related to the Q values of the modules. Indeed, we found that for the 44 skeletons the Q value of a skeleton is well correlated with the product of the Q values of the three modules that make up the skeleton (Figure 4B).

## Discussion

In summary, our study of the relationship between function and topology revealed certain design principles that may be applicable to a broader class of biological systems. We found that the requirement of functional robustness drastically reduces the choices of viable topology. Similar findings were reported in models of circadian oscillators (Wagner, 2005) and, in a broader sense, protein folding (Li et al, 1996), suggesting that the constraint may be general. The approach and method developed here may be applicable in analyzing other networks and in designing novel functional networks.

### Modularity

In our case, the robust topologies are a set of modular combinations. Here modularity arises from the decomposability of the function into relatively independent sub-functions. Combinations of modules provide a combinatorial variability—each sub-function has a multiple choice of modules. Although only a subset of these combinations is robust, this flexibility may be crucial for the network to evolve and adapt in a wide range of situations (Kirschner & Gerhart, 2005). On the other hand, the fact that each module in the network can be traced back to a simpler sub-function suggests that new and more complex functions can be built from the bottom up via combinations of simpler functional modules. Similar principles have been seen in other biological systems, e.g.

the transcriptional control (Carroll, 2005) and protein interactions (Bhattacharyya et al, 2006), suggesting a hierarchical modular design toward an increasing complexity.

**Optimality and pleiotropy**

Another insight gained from our study is that the topology adopted by nature may not necessarily be the most robust *per se*, but may nonetheless be optimized within certain biological constraints. Here the constraint seems to be that no direct positive loops can be used on the three nodes: "E", "S", and "W". Direct positive autoregulation may result in a less flexible system, which may impair the other functional abilities of the Hh and the Wg pathways. Given the multiple tasks carried out by the two major signaling pathways (Galis et al, 2002), it is plausible that when a positive loop is needed for a specific function it is best done with another mediator (here "S") that is only involved in that function. Intriguingly, in the segment polarity gene network, the "S" node is part of the positive loops of both "E" and "W" (Figure 1C and D). The "S" loops with "E" through mutual repression and with "W" through mutual activation. This design ensures that "E" and "W" cannot be switched on in the same cell. Our study suggests that modular design not only provides robustness but can also facilitate variability to accommodate a variety of pleiotropic constraints.

**Evolution**

The distribution of Q values (Figure 2) may have quantitative implications in the early history of evolution. One may ask whether nature picked a robust topology in the first place or a fragile one and then improved upon it. The argument for the former is that a robust topology has a very large working parameter space and thus is easy to be "hit" by random parameter sets. The argument for the latter is that although each particular fragile topology has a tiny working parameter space and is hard to be "hit" there are so many of them that the chance of hitting any is high. This question can be phrased quantitatively by asking what is the most probable Q value for the quantity $Q \times P(Q)$, where $P(Q)$ is the probability density of the Q distribution (inset of Figure 2). We found that $P(Q) \approx c/Q^{\alpha}$, with $\alpha=1.37$, which implies that $Q \times P(Q) \approx c/Q^{0.37}$ has larger weights in smaller Q's, favoring the fragile topologies as nature's first pick.

## Functional versus robustness constraint

We have sampled each of the 14,348,907 3-node topologies with N=100 random sets of parameters. We found that there are M=156,016 (about 1%) networks that are "functional" with at least one parameter set. Among these "functional networks", 96% of them contain at least one module of each kind. Thus it appears that the function alone (without robustness) is a primary constraint on topology. However, note that the number of the "functional networks" M can increase with the sampling number N. We have sampled all 2-node networks with N=100, 1000, and 10000 (Supplementary Information) and found that M=75, 100, and 120, respectively. Furthermore, we found that the percentage of the "functional networks" that are modular (containing at least one skeleton) decreases with N: it is 92% for N=100, 74% for N=1000, and 63% for N=10000. Thus, "functional network" can not be defined unambiguously without a minimal robustness (Q) requirement. There would be more and more non-modular "functional networks" if we sample the parameter space more and more thoroughly. These networks "function" with some special arrangements of parameters. On the other hand, if we focus on robust functional networks (the ones with Q larger than a minimal value), all the statistical properties converge with the sampling number and the conclusions are robust.

## Methods

### The ODE model

For a fixed topology, every cell has the same set of nodes and links. Each node A has a half life time $\tau_A$. Each link is modeled with a Hill function. "A link from A" has either the form $A^n/(A^n+k^n)$ (positive regulation) or $k^n/(A^n+k^n)$ (negative regulation). After proper normalization, each node has one parameter (half life time) and each link has two parameters ($n$ and $k$). Multiple regulations to the same node are modeled as the product of the regulations. For example, for the topology of Figure 5 the equations in each cell are

$$\frac{dA}{dt} = \frac{1}{\tau_A}\left(\frac{k_1^{n1}}{B^{n1}+k_1^{n1}} - A\right)$$

$$\frac{dB}{dt} = \frac{1}{\tau_B}\left(\frac{k_2^{n2}}{A^{n2}+k_2^{n2}} \frac{A_{out}^{n3}}{A_{out}^{n3}+k_3^{n3}} - B\right)$$

where $A_{out}$ is the average concentration of $A$ in the neighboring cells. We use the multiplication rule to model multiple regulations because in the full biological network (Figure 1A and C) the negative links are dominant, implying an "AND"-like logic when a negative link appears together with other regulations. In the simplified model, a positive link can be a result of two negative links (e.g., the link S→W in Figure 1D is from two negatives: Slp—|Mid—|Wg in Figure 1C). In this case, the positive link should also have the "AND"-like logic. For simplicity, we implement the multiplication rule uniformly whenever there are multiple regulations. In our case, we have tested that the simplified models (Figure 1B and D) have the same steady state pattern as the full models (Figure 1A and C).

## Simulation

We use the GNU Scientific Library (GSL) for ODE simulation (Galassi et al, 2002). The function used for the integration is rkf45. Calculation time is set to 800 mins (virtual simulation time). In most calculations, we randomly sample 100~10,000 parameter sets using the LHS method (McKay et al, 1979) which minimizes the correlation between different parameter dimensions. The ranges of the parameters used in the sampling are as follows: $k$=(0.001-1), $n$=(2-10), and $\tau$=(5min-100min). They are similar to the ranges used in previous studies (von Dassow et al, 2000; Ingolia, 2004). $k$ is evenly sampled on the log scale and both $\tau$ and $n$ are evenly sampled on the linear scale. The ODEs are simulated on an 8-cell segment (one row of the parasegment in Figure 1F). Periodic boundary condition is used in both directions ($x$ and $y$).

## Patterning function and judgment of pattern

We judge whether or not a network with a given parameter set can perform the required patterning function in the following way. Let $x(I;n)$ be the value of node $I$ in cell $n$. $I$ can be E, S or W and $x$ is a real number between 0 and 1. The patterning function is defined as: Given the initial condition (Figure 1E) $x(E;1,2)=1$, $x(E;3-8)=0$, $x(S;1-4)=0$, $x(S;5-8)=1$, $x(W;1-6)=0$, $x(W;7,8)=1$, the network should reach the target steady state (Figure 1F) $x(E;1)=1$, $x(E;2-8)=0$, $x(W;1-7)=0$, $x(W;8)=1$ within a given time. We use a similar criterion as the one in previous studies (von Dassow et al, 2000; Ingolia, 2004) to judge if a pattern is acceptable to be the target pattern. Specifically, for node $I$ in cell $n$, a score $T$ is given to evaluate if its expression level is consistent with the target pattern.

$$T_{off} = \alpha_{max} f(x(I,n)) = \alpha_{max} \frac{(x(I,n)/x_t)^3}{1+(x(I,n)/x_t)^3}$$
$$T_{on} = \alpha_{max}(1 - f(x(I,n)))$$

where $x(I,n)$ is the concentration of node $I$ in cell $n$, $x_t$ the threshold for $x$ (we use 10% here), $\alpha_{max}$ the worst-possible score (0.5 here). $T_{off}$ is used when the target state requires that node $I$ has a low value (0) in cell $n$. $T_{on}$ is used when node $I$ should have a high value (1) in cell $n$. All the individual scores are combined to give the total score:

$$\sum_{node}\sum_{cell} T(x(I,n)).$$

If the total score is lower than 0.0125, the pattern is acceptable. This threshold is more stringent than that in the previous work (von Dassow et al, 2000; Ingolia, 2004). We check the pattern twice, at 600 min and at 800 min. If the score is smaller than 0.0125 at both times, we accept the pattern.

## Acknowledgments

We thank Nicholas Ingolia, Morten Kloster, Edo Kussell, Patrick O'Farrell, Wendell Lim, Andrew Murray, Leslie Spector, and members in the Center for Theoretical Biology at PKU for discussion, comments, and/or critical reading of the manuscript. This work was supported by National Key Basic Research Project of China (2003CB715900) and National Natural Science Foundation of China. C.T. acknowledges support from the Sandler Family Supporting Foundation.

# References

Albert R & Othmer HG (2003) The topology of the regulatory interactions predicts the expression pattern of the segment polarity genes in *Drosophila* melanogaster. *J. Theor. Biol.* **223**: 1-18.

Alexandre C, Jacinto A & Ingham PW (1996) Transcriptional activation of hedgehog target genes in *Drosophila* is mediated directly by the cubitus interruptus protein, a member of the GLI family of zinc finger DNA-binding proteins. *Genes Dev* **10**: 2003-2013.

Barkai N & Leibler S. (1997) Robustness in simple biomedical networks. *Nature* **387**: 913-917.

Bhattacharyya RP, Reményi A, Yeh BJ & Lim WA (2006) Domains, Motifs, and Scaffolds: The Role of Modular Interactions in the Evolution and Wiring of Cell Signaling Circuits. *Annu. Rev. Biochem.* **75**: 655-680.

Buescher M, Svendsen PC, Tio M, Miskolczi-McCallum C, Tear G, Brook WG & Chia W (2004) Drosophila T box proteins break the symmetry of hedgehog-dependent activation of wingless. *Curr. Biol.* **14**: 1694-1702.

Carroll SB (2005) Evolution at Two Levels: On Genes and Form. *PLoS Biol.* **3**: e245.

DiNardo S, Heemskerk J, Dougan S & O'Farrell PH (1994) The making of a maggot: patterning the *Drosophila* embryonic epidermis. *Curr. Opin. Genet. Dev.* **4**: 529-534.

DiNardo S, Sher E, Heemskerk-Jongens J, Kassis JA & O'Farrell PH (1988) Two-tiered regulation of spatially patterned engrailed gene expression during *Drosophila* embryogenesis. *Nature* **332**: 604-609.

El-Samad H, Kurata H, Doyle JC, Gross CA & Khammash M (2005) Surviving heat shock: control strategies for robustness and performance. *Proc. Natl. Acad. Sci. USA* **102**: 2736–2741.

Galassi M, Davies J, Theiler J, Gough B, Jungman G, Booth M, Rossi F. (2002) *GNU Scientific Library Reference Manual (2nd Ed.)*, pp. 290-298.

Galis F, van Dooren TJ & Metz JA (2002) Conservation of the segmented germband stage: robustness or pleiotropy? *TRENDS Genet.* **18**: 504-509.

Ingolia NT (2004) Topology and robustness in the *Drosophila* segment polarity network. *PLoS Biol.* **2**: e123.

Kirschner M & Gerhart J (2005) *The Plausibility of Life* (Yale University Press, New Haven and London).

Klingensmith J & Nusse R (1994) Signaling by wingless in *Drosophila*. *Dev. Biol.* **166**: 396-414.

Kollmann M, Lovdok L, Bartholome K, Timmer J & Sourjik V (2005) Design principles of a bacterial signalling network. *Nature* **438**: 504-507.

Lee HH & Frasch M (2000) Wingless effects mesoderm patterning and ectoderm segmentation events via induction of its downstream target sloppy paired. *Development* **127**: 5497-5508.

Li F, Long T, Lu Y, Ouyang Q & Tang C (2004) The yeast cell-cycle network is robustly designed. *Proc. Natl. Acad. Sci. USA* **101**: 4781-4786.

Li H, Helling R, Tang C & Wingreen N (1996) Emergence of preferred structures in a simple model of protein folding. *Science* **273**: 666-669.

Lum L & Beachy PA (2004) The Hedgehog response network: sensors, switches, and routers. *Science* **304**: 1755-1759.

Martizez Arias A (1993) in *Development and patterning of the larval epidermis of Drosophila,* eds. Bate M & Hartenstein V (Cold Spring Harbor Laboratory Press, Cold Spring Harbor, NY), pp. 517-608.

McKay MD, Beckman RJ & Conover WJ (1979) A comparison of three methods for selecting values of input variables in the analysis of output from a computer code. *Technometrics* **21**: 239-245.

Meir E, Munro EM, Odell GM & von Dassow G (2002) Ingeneue: A versatile tool for reconstituting genetic networks, with examples from the segment polarity network. *J Exp Zool* **294**: 216-251

Milo R, Shen-Orr S, Itzkovitz S, Kashtan N, Chklovski D & Alon U (2002) Network motifs: simple building blocks of complex networks. *Science* **298**: 824-827.

Peel AD, Chipman AD & Akam M (2005) Arthropod segmentation: beyond the *Drosophila* paradigm. *Nat. Rev. Genet.* **6**: 905-916.

Perrimon N (1994) The genetic basis of patterned baldness in *Drosophila*. *Cell* **76**: 781-784.

Prill RJ, Iglesias PA & Levchenko A (2005) Dynamic Properties of Network Motifs Contribute to Biological Network Organization. *PLoS Biol.* **3**: e343.

Raff RA (1996) *The Shape of Life* (University of Chicago Press, Chicago).

Sander K (1983) in *Development and Evolution,* eds. Goodwin, B. C., Holder, N. & Wylie, C. C. (Cambridge University Press, Cambridge), pp. 137–154.

Schuster P, Fontana W, Stadler PF & Hofacker I (1994) From sequences to shapes and back: a case study in RNA secondary structures. *Proc. Roy. Soc.* (London) **B 255**:279-284.

Siegfried E, Wilder EL & Perrimon N (1994) Components of wingless signalling in *Drosophila*. *Nature* **367**: 76-80.

Tabata T, Eaton S & Kornberg TB (1992) The *Drosophila* hedgehog gene is expressed specifically in posterior compartment cells and is a target of engrailed regulation. *Genes Dev* **6**: 2635-2645.

Vincent JP & O'Farrell PH (1992) The state of engrailed expression is not clonally transmitted during early *Drosophila* development. *Cell* **68**: 923-931.

von Dassow G, Meir E, Munro EM & Odell GM (2000) The segment polarity network is a robust developmental module. *Nature* **406**: 188-192.

von Dassow G & Odell GM (2002) Design and constraints of the *Drosophila* segment polarity module: robust spatial patterning emerges from intertwined cell state switches. *J. Exp. Zool.* **294**: 179-215.

Waddinton CH. (1957) *The Strategy of the Genes.* London: George Allen & Unwin Ltd.

Wagner A (2005) Circuit topology and the evolution of robustness in two-gene circadian oscillators. *Proc. Natl. Acad. Sci. USA* **102**: 11775-11780.

Wolpert L, Beddington R, Jessell T, Lawrence P, Meyerowitz E & Smith J (2002) *Principles of Development* (Oxford Press, New York).

# Figure Legends

**Figure 1.** Segment polarity network and expression pattern of *wg* and *en*. (*A*) The segment polarity gene network model of (Ingolia 2004). Ellipses represent mRNAs and rectangles proteins. Lines ending with an arrow and a dot denote activation and repression, respectively. Dashed lines indicate intercellular regulations. The grey line means no direct biological evidence. Nodes are colored into three groups, each of which is represented by one node in (B). (*B*) The simplified topology of (A). Each node here represents a group of nodes in (A) of the same color. (*C*) Our model of the segment polarity gene network (see also (von Dassow & Odell 2002)). Slp regulates *wg* positively through the *mid* gene and its product, which is represented by an arrow from "S" to "W" in (D). (*D*) The simplified topology of (C). (*E*) The initial condition of the patterning function. In 3-node networks, "S" expresses in the posterior 4 cells of the parasegment. The pattern is periodic. (*F*) The final stable pattern. In 3-node networks, "S" is not fixed to be any specific pattern in the final state. (*G*) *zw3* mutant phenotype. (*H*) *ptc* mutant phenotype. Note that E-F is a simple representation of the actual embryo surface which is extended in both directions and includes 14 segments.

**Figure 2.** The histogram of Q values for 3-node networks. Each of the 14,348,907 networks is sampled with 100 random parameter sets (black bars). Each of the resulting 156,016 networks with Q>0 is resampled with 1000 random parameter sets (only data with Q>0.1 are shown; red bars). Inset: the same data plotted in log-log scale and in terms of the probability density. The straight line has a slope of -1.37.

**Figure 3.** Skeletons and Functional modules. (*A*) The four skeletons in robust 2-node topologies (black lines). The green, orange and red links are neutral, bad and very bad links, respectively. The numbers below the skeletons are (its Q-value, the size of its family). (*B*) The three kinds of modules correspond to the three sub-functions in 3-node networks. The bold modules are also those of the 2-node networks. Many of these modules can be identified as significant network motifs among all networks with Q>0.1 (see details in supporting information). The combination of these modules leads to 44 robust core topologies or skeletons. The number under each module is (its Q-value, the frequency the module is being used in the 44 skeletons).

**Figure 4.** The number of networks in a family (*A*) and the skeleton's estimated Q value (the product of the modules' Q values) (*B*) versus the Q value of the skeleton.

**Figure 5.** An example of a 2-node topology. Intracellular regulations (solid lines) act on nodes within the cell; intercellular regulations (dashed lines) act on target nodes in nearby cells.

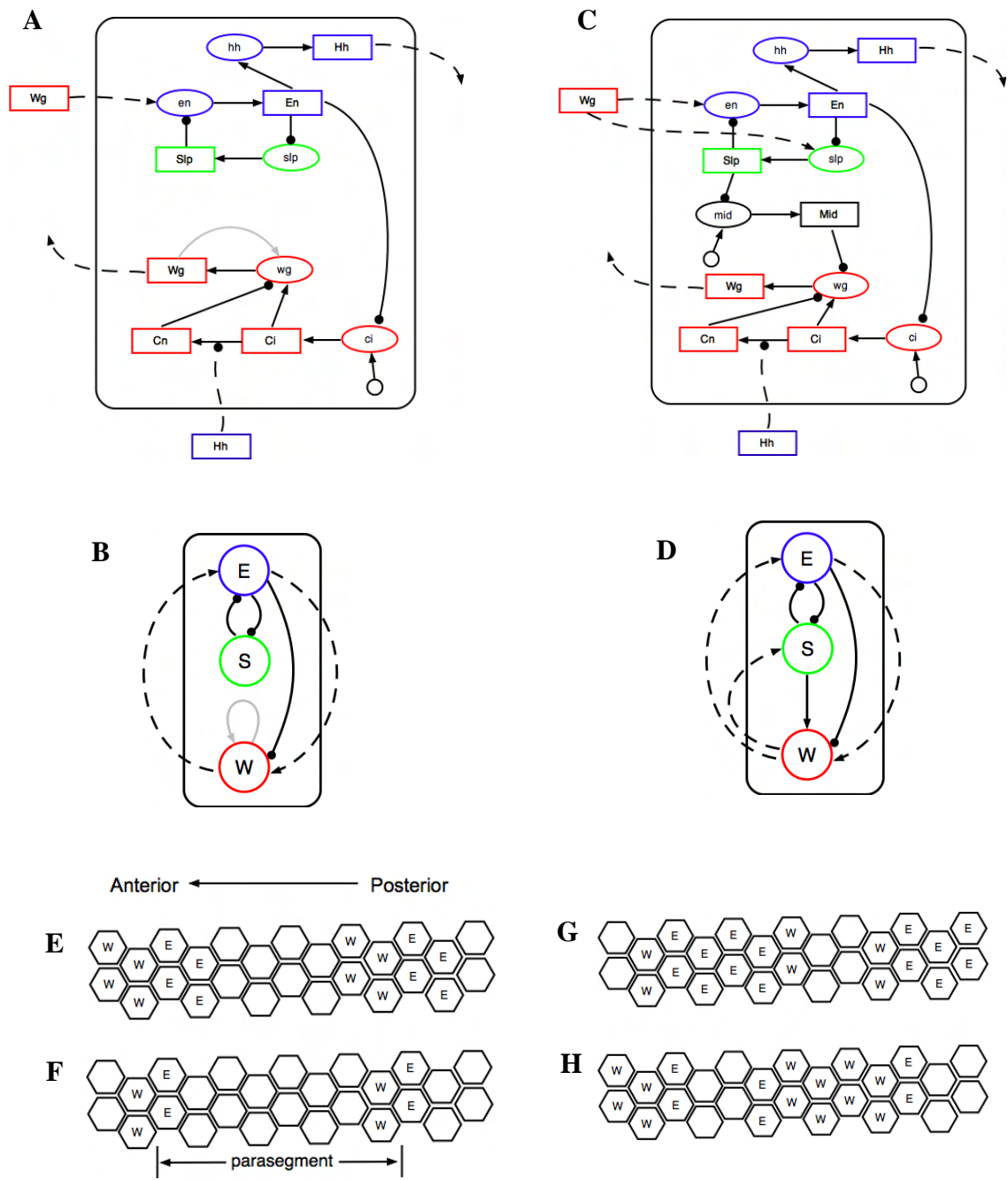

Figure 1

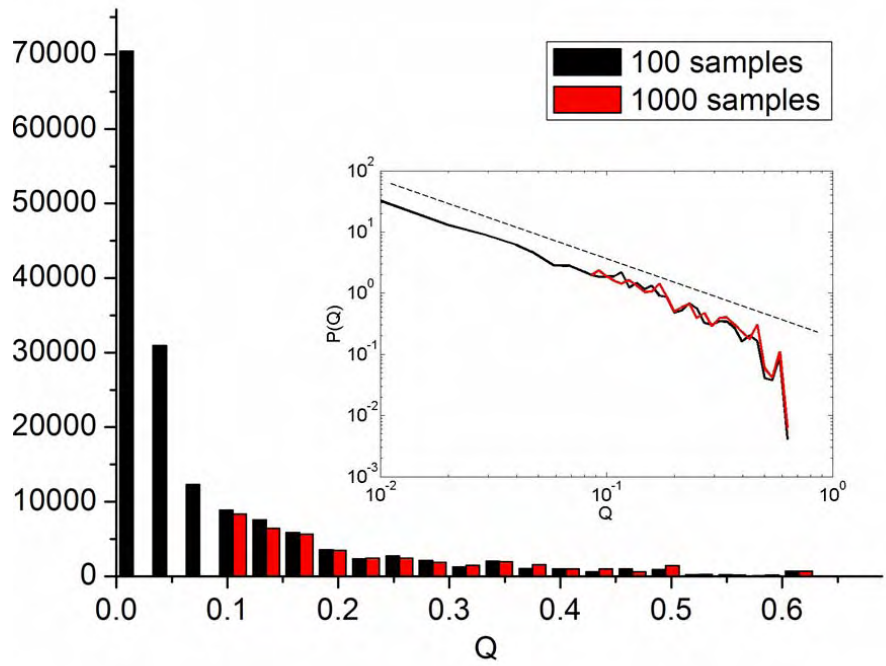

Figure 2

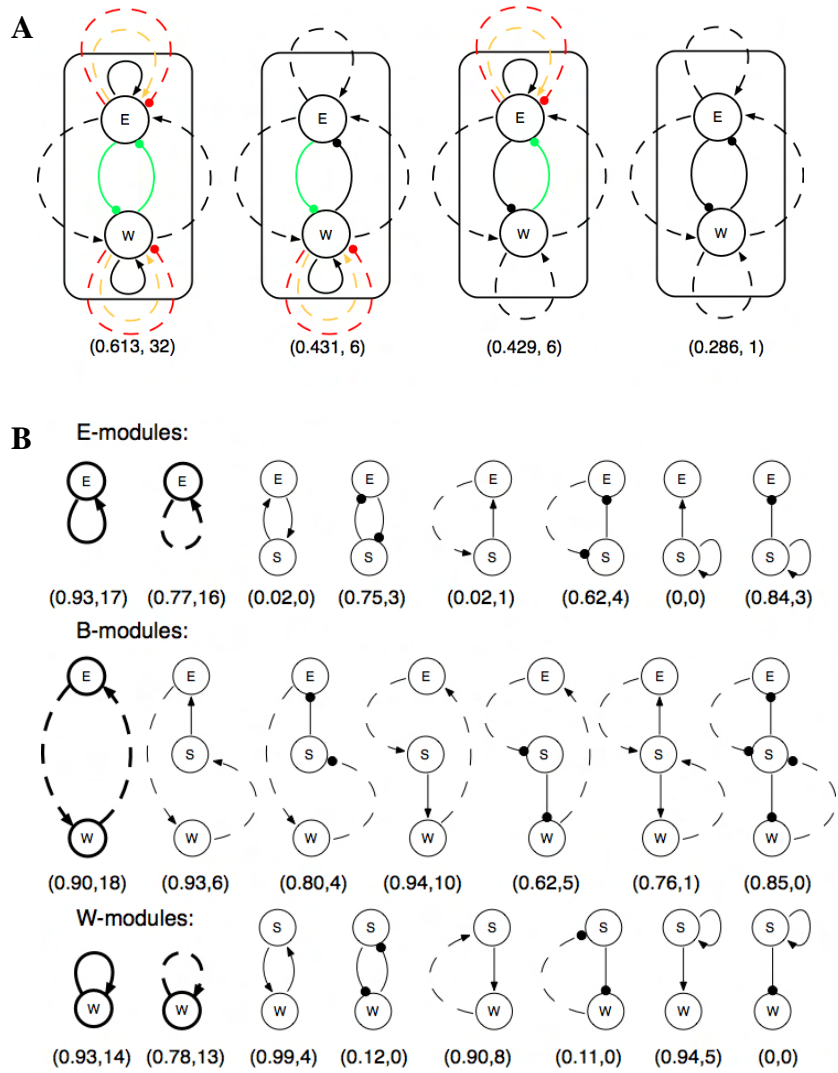

Figure 3

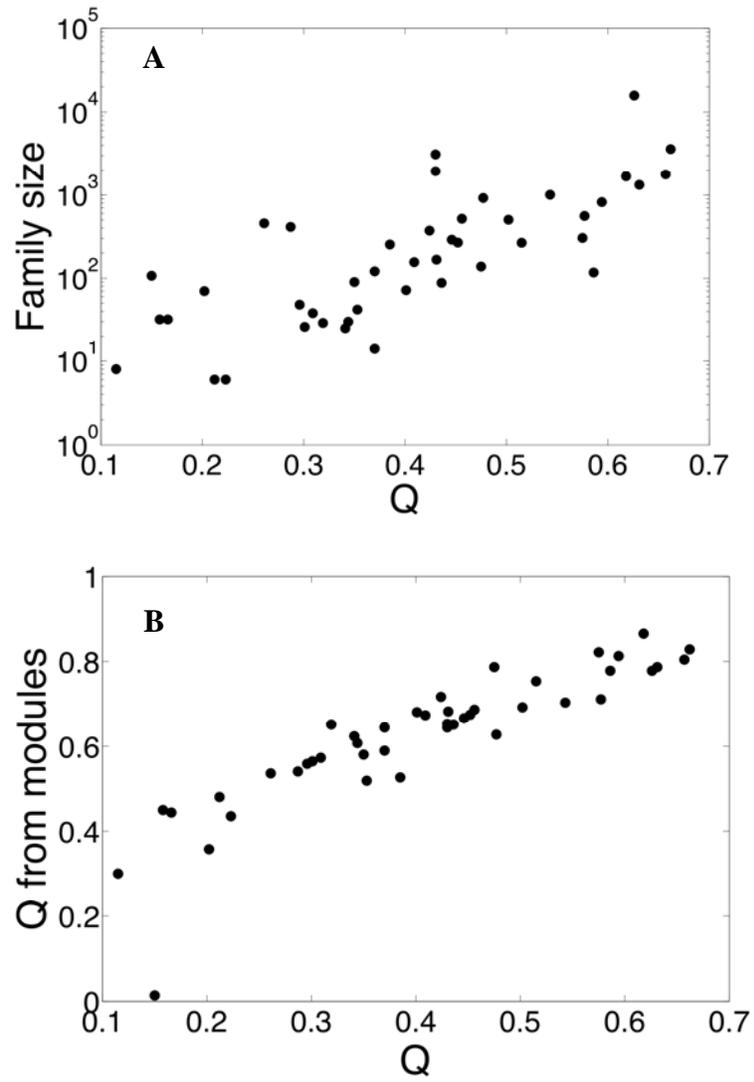

Figure 4

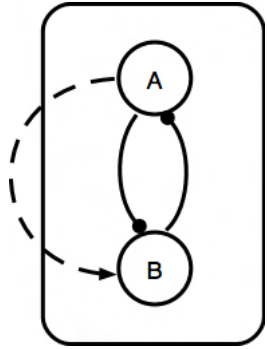

Figure 5